\pdfoutput=1

\documentclass[12pt]{article}

\usepackage{xspace}
\usepackage{rotating,multirow,booktabs}
\usepackage{graphicx}
\usepackage{amssymb}

\def\etal{{\it et al.}\xspace}

\begin{document}

\begin{titlepage}
\begin{center}
\Huge {\bf Comparison of available measurements of the absolute fluorescence yield}
\par
\vspace*{2.0cm} \normalsize {\bf {J. Rosado, F. Blanco, F. Arqueros}}
\par
\vspace*{0.5cm} \small \emph{Departamento de F\'{i}sica At\'{o}mica, Molecular y Nuclear, Facultad de Ciencias F\'{i}sicas,
Universidad Complutense de Madrid, E-28040 Madrid, Spain}
\end{center}
\vspace*{2.0cm}

\begin{abstract}
The uncertainty in the absolute value of the fluorescence yield is still one of the main contributions to the total error in the
reconstruction of the primary energy of ultra-energetic air showers using the fluorescence technique. A significant number of
experimental values of the fluorescence yield have been published in the last years, however reported results are given very
often in different units (photons/MeV or photons/m) and for different wavelength intervals. In this work we present a comparison
of available results normalized to its value in photons/MeV for the 337~nm band at 800~hPa and 293~K. The conversion of
photons/m to photons/MeV requires an accurate determination of the energy deposited by the electrons in the field of view of the
experimental setup. We have calculated the energy deposition for each experiment by means of a detailed Monte Carlo simulation
including when possible the geometrical details of the particular setup. Our predictions on deposited energy, as well as on some
geometrical factors, have been compared with those reported by the authors of the corresponding experiments and possible
corrections to the fluorescence yields are proposed.

\end{abstract}
\end{titlepage}

\section{Introduction}
\label{sec:intro}

Charged particles of a cosmic ray shower, mostly electrons, passing through the atmosphere lose energy by inelastic collisions
with air molecules. A small fraction of the total deposited energy is emitted by molecular nitrogen as UV fluorescence radiation
in the $\sim 300 - 400$~nm spectral range. The detection of this radiation provides a precise determination of the longitudinal
profile allowing the reconstruction of the cosmic ray properties, in particular its energy. This technique~\cite{Bunner_PhD} was
first successfully used by the Fly's Eye experiment~\cite{FlysEye} and later by HiRes~\cite{HiRes}. Fluorescence telescopes are
presently being employed by the Pierre Auger Observatory~\cite{Auger} and the Telescope Array experiment~\cite{TelescopeArray}.
On the other hand, the JEM-EUSO project~\cite{JEM_EUSO} and the S-EUSO free-flying satellite mission~\cite{S_EUSO} are being
designed for the detection of fluorescence traces of air showers viewing downward from the top of the atmosphere.

The ratio between number of fluorescence photons and deposited energy, i.e., the fluorescence yield $Y$, is a key calibration
parameter determining the energy scale of the fluorescence telescopes. A number of absolute measurements of the fluorescence
yield have been carried out in laboratory experiments in the last years.

At atmospheric pressure, fluorescence emission in the spectral range of interest basically comes from the Second Positive (2P)
system of N$_2$ and, to a much less extent, the First Negative (1N) system of N$^+_2$. Each one results from the transition
between two electronic states and consists of a set of molecular bands corresponding to different combinations of vibrational
levels $v$-$v'$ of the respective upper and lower states. Excitation cross section of the 1N system decreases with energy in the
$10^2-10^6$~eV range showing a very smooth growing behavior at larger energies. That of the 2P system peaks at about 15~eV
decreasing strongly with an $E^{-2}$ dependence. As a consequence, high-energy electrons themselves are very inefficient for the
generation of air fluorescence. In fact, as is well known, fluorescence emission along the track of an energetic primary
electron is mainly induced by low-energy secondary electrons ejected in successive ionization processes. The efficiency for
fluorescence emission varies with the energy spectrum of these secondaries\footnote{For instance, electrons with energies below
the threshold for fluorescence excitation promptly lose all their energy, but induce no fluorescence.}, which in principle
depends on the primary energy. However, a theoretical calculation using a detailed Monte Carlo
simulation~\cite{Blanco,Arqueros_Astropart_Phys,Arqueros_NJP} has demonstrated that the fluorescence yield is nearly constant
for energies above a few kiloelectronvolts. This $E$-independent behavior of the fluorescence yield has been checked
experimentally~\cite{FLASH_thick,MACFLY,5th_FW_AIRFLY_E}.

A number of experimental values of the absolute fluorescence yield are available in the literature. Essentially, in these
experiments a beam of electrons crosses a collision chamber filled with air at known conditions. However different techniques
can be used. Many authors~\cite{MACFLY,Kakimoto,Nagano_03,Nagano_04,Lefeuvre,AirLight} have used electrons from a source of
$^{90}$Sr with energy around 1~MeV. Other absolute measurements have been performed with higher energy electrons from
accelerators. The MACFLY Collaboration~\cite{MACFLY} used the CERN/SPS-X5 test beam facility which delivers a pulsed electron
beam of about 10$^4$ electrons per spill (4.8~s duration) every 16.8~s. Measurements at 20 and 50~GeV were carried out using
this facility. The FLASH Collaboration~\cite{FLASH_06,FLASH_08} used the Final Focus Test Beam facility at the Stanford Linear
Accelerator Center which provided 28.5~GeV electrons in 3~ps pulses of about 10$^8$ electrons at a rate of 10~Hz. The AIRFLY
Collaboration is carrying out absolute measurements of the air-fluorescence yield using different accelerators. Preliminary
results obtained with 350~MeV electrons at the Beam Test Facility of the Instituto Nazionale di Fisica Nucleare have been
presented in~\cite{5th_FW_AIRFLY}. New measurements using a 120~GeV proton beam from the Meson Beam Test Facility at Fermilab
were presented at the 6th AFW~\cite{6th_FW} and very likely final results will be published soon.

In this work we will compare the absolute values reported by the above authors. In many occasions comparison cannot be done
directly. For instance, some authors measure single intense fluorescence bands while others detect the integrated fluorescence
in a wide spectral range. Comparison between these experimental results can be carried out if the relative intensities of the
bands along the fluorescence spectrum are known. We will normalize all the results to the most intense fluorescence band at
337~nm using accurate experimental data in full agreement with theoretical intensity ratios proposed
in~\cite{Blanco,Arqueros_Astropart_Phys,Arqueros_NJP}.

On the other hand, some authors report their measurements in units of photons per electron and meter while others give directly
the $Y$ parameter in photons per unit deposited energy (e.g., photons/MeV). The conversion factor between both magnitudes is the
energy deposited per electron and unit path length. As it has been pointed
out~\cite{Blanco,Arqueros_Astropart_Phys,Arqueros_NJP}, an accurate determination of the deposited energy requires discounting
that corresponding to secondary electrons outside the field of view of the optical system used to detect the emitted
fluorescence. To evaluate this effect we have carried out a MC simulation of the various experimental setups. When possible a
dedicated simulation including geometrical details has been performed. Also the effect of the spatial distribution of deposited
energy on the optical efficiency of the experimental setup has been evaluated in some cases. As a result of this study absolute
values of the fluorescence yield normalized to that of the 337~nm band at 293~K and 800~hPa in dry air will be shown with a
discussion on possible corrections to be applied. A previous comparison at 1013~hPa for some of the above measurements has been
published recently~\cite{Arqueros_NJP}. In this work we will update those results and extend the comparison to other
experimental results.

\section{Fluorescence Yield}
\label{sec:fluorescence_yield}

\subsection{Units}
\label{sec:units}

In the literature, several parameters have been used to measure the amount of fluorescence light induced by an electron moving
in the atmosphere. A detailed description of the various magnitudes can be found in~\cite{5th_FW_summary}. The number of photons
emitted per electron and unit path length have been used extensively. We will name this parameter as $\varepsilon$ and, for a
band $v-v'$, it is proportional to the number of nitrogen molecules per unit volume $N$ and the reciprocal of the $1+P/P'_v$
Stern-Volmer factor, accounting for the collisional de-excitation of the upper level $v$, where $P'_v$ is a characteristic
pressure depending on the gas composition and temperature. If ignoring the contribution of secondary electrons, the
proportionality constant would be the so-called optical cross section.

While most secondary electrons have low energy, and thus, short range at atmospheric pressure, a small number of high-energy
secondaries can escape the observation region. As a consequence, in a laboratory experiment a fraction of the fluorescence light
cannot be detected. Following the parametrization in~\cite{Blanco,Arqueros_Astropart_Phys,Arqueros_NJP}, an effective optical
cross section $\sigma_{vv'}^{\rm eff}$ can be defined in such a way that the total number of photons of the $v-v'$ band detected
under given experimental conditions (including contributions from both primary and secondary electrons) can be expressed as

\begin{equation}\label{eq:varepsilon}
\varepsilon_{vv'}=N\frac{\sigma^{\rm eff}_{vv'}}{1+P/P'_v}\,.
\end{equation}

The parameter $\sigma_{vv'}^{\rm eff}$ depends on $E$, $P$ and the geometrical features of the experimental setup. Therefore,
even for the same primary energy, the measured $\varepsilon_{vv'}$ value depends on the particular experiment. In addition, the
pressure dependence of $\varepsilon_{vv'}$ as would be predicted by the Stern-Volmer law is distorted in a geometry-dependent
way.

A further pressure dependence arises from the vibrational relaxation of excited nitrogen induced by collisions with surrounding
nitrogen molecules in the ground state~\cite{Dilecce}. The effect of vibrational relaxation is twofold. Firstly, it contributes
to the total quenching of a given level $v$ and thus is included in the Stern-Volmer factor~(\ref{eq:varepsilon}). Secondly,
vibrational relaxation of upper levels ($>v$) provides an additional excitation channel. Although, in principle, this effect may
lead to a departure from the Stern-Volmer law, significant deviations have not been appreciated experimentally for any N$_2$
band in air~\cite{AIRFLY_P}. In fact, for the 2P(0,0) band, for which detailed data on vibrational quenching are
available~\cite{Dilecce}, a simple calculation shows that the only relevant effect of vibrational quenching is a variation of
the resulting characteristic pressure respect to the one from pure electronic quenching~\cite{Morozov}.

The number of photons per unit energy deposited in the medium is a more appropriate parameter for use in fluorescence detectors
of cosmic ray showers. Following an already common usage, we will keep the term fluorescence yield to refer to this magnitude
$Y$. For a given band, $\varepsilon_{vv'}$ and $Y_{vv'}$ are related by

\begin{equation}\label{eq:Y}
Y_{vv'}=\frac{\varepsilon_{vv'}}{({\rm d}E/{\rm d}x)_{\rm dep}}\,,
\end{equation}
where $({\rm d}E/{\rm d}x)_{\rm dep}$ is the energy deposited per primary electron and unit path length. In the above equation,
both the emitted photons and the deposited energy should correspond to the same volume, which is defined by the field of view of
the experimental detection system. Note that a significant fraction of energy is deposited by secondary electrons which move
away from the path of the primary electron. It has been shown~\cite{Arqueros_Astropart_Phys,Arqueros_NJP} that
$N\sigma_{vv'}^{\rm eff}$ and $({\rm d}E/{\rm d}x)_{\rm dep}$ are proportional, resulting $Y_{vv'}$ to be given by

\begin{equation}\label{eq:Y0}
Y_{vv'} = \frac{Y^0_{vv'}}{1+P/P'_{vv'}}\,,
\end{equation}
where the $Y^0_{vv'}$ parameter is independent of the experimental conditions and represents the fluorescence yield at null
pressure, i.e., in the absence of quenching.

While measurements of the fluorescence yield at atmospheric pressure by different authors show a reasonable agreement, the
available data on the $P'_v$ parameters show large discrepancies\footnote{Notice that many authors ignore the additional
pressure dependence of $\varepsilon_{vv'}$ due to secondary electrons~\cite{Arqueros_NJP}.}. Therefore the extrapolation of the
fluorescence yield to null pressure is unsafe, making the $Y^0_{vv'}$ parameter unsuitable for our proposes. In this work,
comparison will be thus performed using the fluorescence yields measured at given atmospheric conditions.

\subsection{Deposited energy}
\label{sec:Edep}

The energy deposited by an electron per unit path length within a given volume can be expressed~\cite{Arqueros_NJP} as

\begin{equation}\label{eq:Edep}
\left(\frac{{\rm d}E}{{\rm d}x}\right)_{\rm dep}=N_{\rm air}\left[\langle E^0_{\rm dep}\rangle+\langle E_{\rm
dep}\rangle\right]\sigma_{\rm ion}\,,
\end{equation}
where $N_{\rm air}$ is the density of air molecules and $\sigma_{\rm ion}$ is the ionization cross section, including the
density correction for high energies. The parameters $\langle E^0_{\rm dep}\rangle$ and $\langle E_{\rm dep}\rangle$ are the
average values of the energy deposited per ionization process within the observation volume by the primary electron itself and
the secondaries, respectively. For electron energies above 410~eV, contribution from X-rays arising from K-shell ionization
should be also added to the energy deposition~\cite{Arqueros_NJP}.

The $\langle E^0_{\rm dep}\rangle$ parameter can be calculated from

\begin{equation}\label{eq:E0dep}
\langle E^0_{\rm dep}\rangle=\langle E_{\rm exc}\rangle\frac{\sigma_{\rm exc}}{\sigma_{\rm ion}}+\langle E^{\rm ion}_{\rm
exc}\rangle+I\,,
\end{equation}
where $\langle E^{\rm ion}_{\rm exc}\rangle$ and $I$ are the average excitation energy and the ionization potential of the
ionized molecule, respectively, both contributing to the energy deposition of the primary electron after an ionizing
interaction. The first term in the right hand of the equation accounts for the contribution from direct excitations without
ionization, being $\langle E_{\rm exc}\rangle$ the average excitation energy of the molecule and $\sigma_{\rm exc}$ the
excitation cross section for a primary electron.

The parameters involved in the above expressions are either available in the literature or can be inferred from molecular data,
except for $\langle E_{\rm dep}\rangle$, which has to be calculated from a simulation of the secondary processes inside the
observation volume.

Next we will describe a Monte Carlo algorithm to calculate both the energy deposited by secondary electrons and their
contribution to the fluorescence emission for a simple geometry. In section~\ref{sec:detailed_simulation} a more detailed
simulation including the geometrical features of several experiments will be presented.

\subsection[Simple geometry]{Monte Carlo analysis under simple geometrical \\assumptions}
\label{sec:simple_geometry}

Ignoring fine geometrical details, the observation region of an experimental setup can be described by a sphere of radius $R$.
In addition, we will assume that primary electrons interact at its center from where secondaries are ejected and tracked until
they escape the sphere or are stopped. A Monte Carlo algorithm for the calculation of both deposited energy and fluorescence
emission in this simple geometry has been described in detail in~\cite{Arqueros_Astropart_Phys,Arqueros_NJP}. Next the main
features and ingredients of the simulation will be outlined and some improvements described.

The algorithm generates secondary electrons with energies distributed as given by the extended Opal formula proposed
in~\cite{Arqueros_NJP}, which depends on the input primary energy. Each generated electron takes a step of random size with mean
value $(N_{\rm air}\sigma)^{-1}$, where $\sigma$ is the total cross section for the electron energy. After this step, the type
of process is randomly chosen between elastic, excitation, ionization and bremsstrahlung, being their relative weights the
corresponding cross sections. While there are no energy losses in elastic interactions, in excitation and ionization processes
the electron is assumed to locally deposit $\langle E_{\rm exc}\rangle$ and $\langle E^{\rm ion}_{\rm exc}\rangle+I$,
respectively. In case of ionization, the electron loses an additional amount of energy to generate a new secondary, whose energy
is randomly calculated from the above-mentioned extended Opal formula. With a probability corresponding to the cross section for
K-shell ionization, also a 410~eV X-ray can be generated at the interaction point, producing another ionization at a random
distance\footnote{In previous versions of our Monte Carlo algorithm~\cite{Arqueros_Astropart_Phys,Arqueros_NJP}, X-rays were not
propagated, assuming all their energy to be locally deposited without producing further ionizations. While this approximation
has no significant effect on the deposited energy, due to the short range of the X-rays, it leads to an underestimation of about
10\% for the total fluorescence emission.} of mean value given by the photon attenuation coefficient from~\cite{NIST_Xrays}. In
case of bremsstrahlung, the energy loss is estimated as 3\% of the electron energy and the produced photons are assumed to
escape the sphere\footnote{We checked that contribution from bremsstrahlung photons to the energy deposition is negligible even
for large experimental setups and GeV electrons, where energy losses due to bremsstrahlung are dominant.}. If an excitation
(ionization) process occurs, the amount of 2P (1N) fluorescence light emitted is calculated from the ratio of the 2P (1N) cross
section to the total excitation (ionization) one. The direction of the electron after an interaction is calculated from an
approximated differential cross section~\cite{Roldan}. The stepped trajectory of each electron is so iteratively simulated until
it escapes the sphere or its energy is below 11~eV (i.e., the threshold for fluorescence production), when the electron is
considered to be stopped and deposits its remainder energy. New secondary electrons arising from ionization processes are added
to the list of electrons to be simulated following the same procedure.

Input parameters in the algorithm are the energy of the primary electron, the density of air molecules (or pressure at given
temperature) and the radius of the sphere. As a result of the simulation the value of $\langle E_{\rm dep}\rangle$ is obtained,
and then, the total energy deposited per primary path length is calculated from expressions~(\ref{eq:Edep})
and~(\ref{eq:E0dep}). Likewise, the $\sigma^{\rm eff}_{vv'}$ parameter can be calculated from the average number of fluorescence
photons emitted inside the sphere in the absence of quenching resulting from the simulation. As expected, both magnitudes turn
out to be a function of the product $R\cdot N_{\rm air}$, that is, proportional to the number of molecules that a secondary
electron encounters in its way out of the sphere, and the quotient provides the $Y^0_{vv'}$ parameter, which is basically
independent of $E$ and $R\cdot N_{\rm air}$.

As discussed in~\cite{Arqueros_NJP}, our calculation of the fluorescence emission is subjected to a systematic uncertainty, that
we estimate in about a 25\%, due to the uncertainties in the various molecular parameters. However the evaluation of the energy
deposited in the sphere is much more accurate. In fact, in the $R\cdot N_{\rm air}\rightarrow\infty$ limit, for which no
secondary escapes, the predictions from the simulation are in perfect agreement with the Bethe-Bloch formula for the collisional
energy loss~\cite{Seltzer} including the corresponding density correction. For this result, a crucial ingredient in the
simulation is the energy distribution of secondary electrons. The extended Opal formula proposed in~\cite{Arqueros_NJP} provides
this spectrum for a very wide energy interval (eV~-~GeV) of both primary and secondary electrons.

Simulation results of the deposited energy for several $R\cdot N_{\rm air}$ values have been presented in~\cite{Arqueros_NJP}.
For the purpose of comparison, in figure~\ref{fig:Edep_comparison} the energy deposited by an electron per unit mass
thickness\footnote{Results expressed in MeV per unit mass thickness are valid for both dry air and pure nitrogen since inelastic
cross sections are approximately proportional to the atomic number $Z$, which is a factor of two lower than the mass number $A$
for both Nitrogen and Oxygen.} in dry air at atmospheric pressure is represented against the primary energy for $R=1$~cm and
10~cm. Below 0.5~MeV the fraction of energy deposited outside the sphere is negligible at this pressure and therefore the energy
deposition for both $R$ values approaches the energy loss. As can be seen in the figure, the deposited energy is very weakly
dependent on the size of the observation region at given pressure. For instance, the energy deposited by electrons of 1~MeV
(1~GeV) increases by 9\% (6\%) when $R$ increases from 1 to 10~cm at atmospheric pressure. This justifies that a reasonable
accuracy in the deposited energy can be achieved even though the fine geometrical details of the experimental setup are not
taken into account.

Notice that in our algorithm, individual electron-molecule collisions as well as the emission of 1N and 2P fluorescence are
simulated with a cutoff energy of 11~eV. On the contrary, usual MC codes (e.g., GEANT and EGS) transports particles in
macroscopic steps adding up the energy deposition in each step. Fluorescence emission, which is mostly generated by electrons
with energies of about a few tens of electronvolts, cannot be calculated by these codes unless the user assumes a certain value
of the fluorescence yield.

\subsubsection[Comparison of deposited energy]{Comparison of results on deposited energy with other \\simulations}
\label{sec:Edep_comparison}

\begin{figure}
\includegraphics[width=\linewidth]{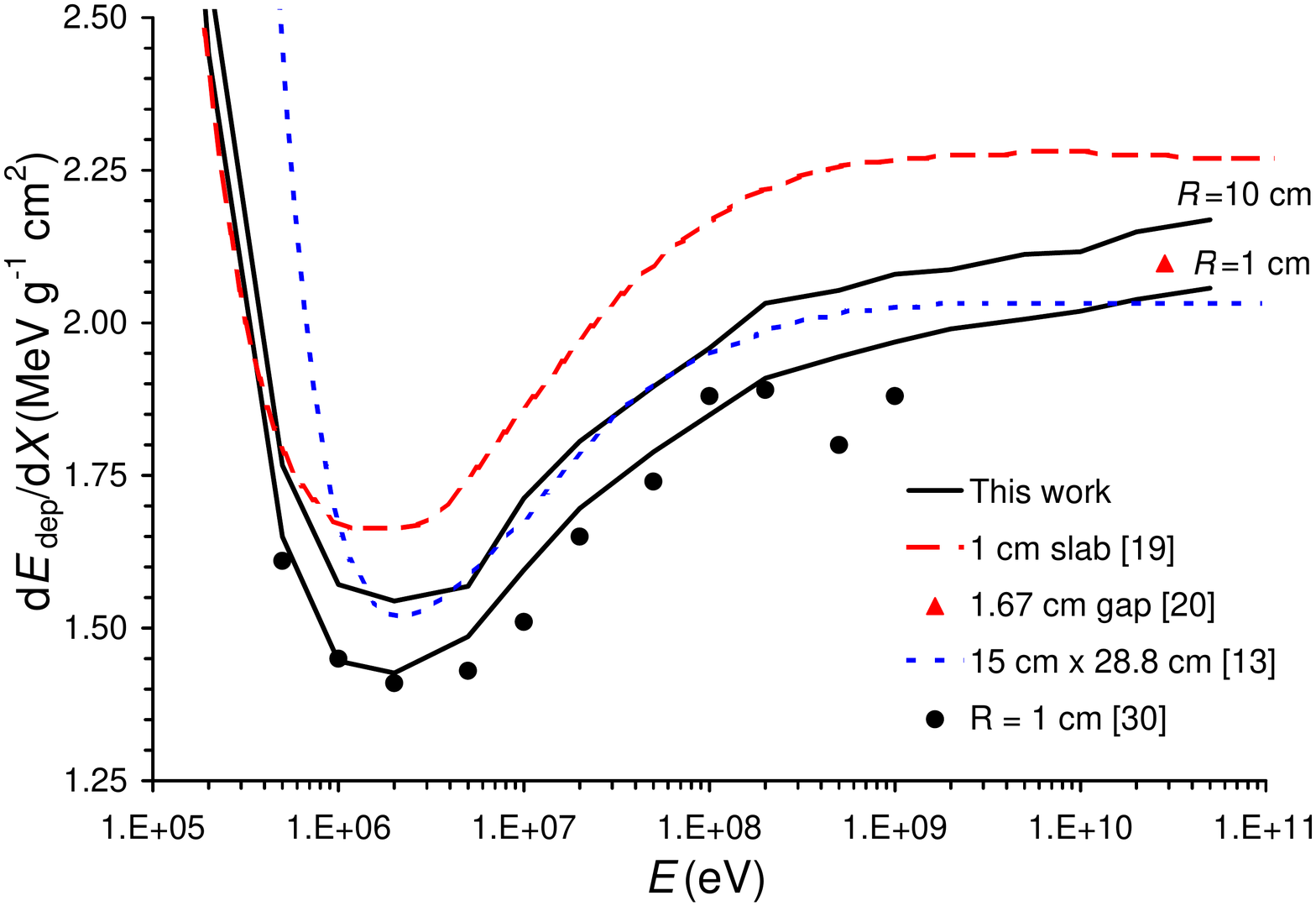}
\centering \caption{Energy deposited per electron and unit mass thickness at atmospheric pressure as a function of primary
energy. Results from our simulation for air spheres of radius $R=1$~cm and $R=10$~cm are shown (solid lines) for comparison with
the energy deposition values reported by other authors (symbols and broken lines). See text for
details.}\label{fig:Edep_comparison}
\end{figure}

The experimental determination of the fluorescence yield requires the calculation of the energy deposited in the observation
volume. Consequently, any possible error in the evaluation of this magnitude translates to the fluorescence yield. Several
experiments have provided results on their simulations which can be compared with our predictions for typical sizes of
experimental setups (see figure~\ref{fig:Edep_comparison}).

In their paper of 2006 the FLASH Collaboration~\cite{FLASH_06} showed results on deposited energy versus $E$ for a 1~cm thick
slab of air at atmospheric pressure obtained with GEANT3, which are significantly larger than our predictions for $R=1$~cm. In a
more recent paper~\cite{FLASH_08}, they obtained an energy deposition for 28.5~GeV electrons crossing a 1.67~cm air gap using
EGS4 which is closer to ours. In next section a detailed comparison is shown.

The MACFLY Collaboration~\cite{MACFLY}, using GEANT4, reported a $({\rm d}E/{\rm d}x)_{\rm dep}$ curve obtained for a
cylindrical chamber, 15~cm in diameter and 28.8~cm long. Its results above 2~MeV are in fair agreement with our predictions for
a sphere of radius 10~cm at atmospheric pressure, except for the highest energies ($>100$~MeV), where our values are up to 6\%
larger. At lower energies the values are oddly much larger than expected (see also the next section for more details).

Finally, the AIRFLY Collaboration, using GEANT4, has provided us with the results of a dedicated simulation for our simple
geometry (sphere of radius 1~cm and primary electrons are forced to interact in its center)~\cite{AIRFLY_per_comm}, which are
fully compatible with ours within statistical fluctuations.

\section[Detailed simulation]{Detailed simulation of fluorescence yield \\experiments}
\label{sec:detailed_simulation}

In this section, a simulation including the particular geometry for several experiments as well as the propagation of beam
electrons will be presented. That allowed the calculation of the energy deposited in the corresponding observation volume. When
possible, the average geometrical acceptance for fluorescence detection has been evaluated and compared with that assumed by the
authors of the experiment. For that purpose, the solid angle subtended by the optical system's entrance pupil is averaged along
the electron trajectories, weighting with the amount of light emitted at each point. Note that the effect on the detection
efficiency due to the spread of electrons in the interaction volume is presumably small, as it results from the average of a
distribution with axial symmetry around the beam axis. That is, the enhanced fluorescence contribution from electrons
approaching the fluorescence detector is nearly compensated by the reduced one from those moving in the opposite direction. In
addition, in those cases where the electron-photon coincidence technique is used (e.g., with a radioactive source), the
detection of beam electrons is required in the simulation, and thus, primaries not firing the detector are rejected. This may
affect the energy spectrum of contributing electrons, and so, the determination of the deposited energy.

We have performed this detailed simulation for the experiments of Nagano \etal~\cite{Nagano_03,Nagano_04},
AirLight~\cite{AirLight}, FLASH-2008~\cite{FLASH_08} and MACFLY~\cite{MACFLY}. Experiments of Kakimoto \etal~\cite{Kakimoto} and
Lefeuvre \etal~\cite{Lefeuvre} will be also discussed although a dedicated simulation was not carried out.

\subsection{Nagano's experiment}
\label{sec:Nagano}

\begin{figure}
\includegraphics[width=\linewidth]{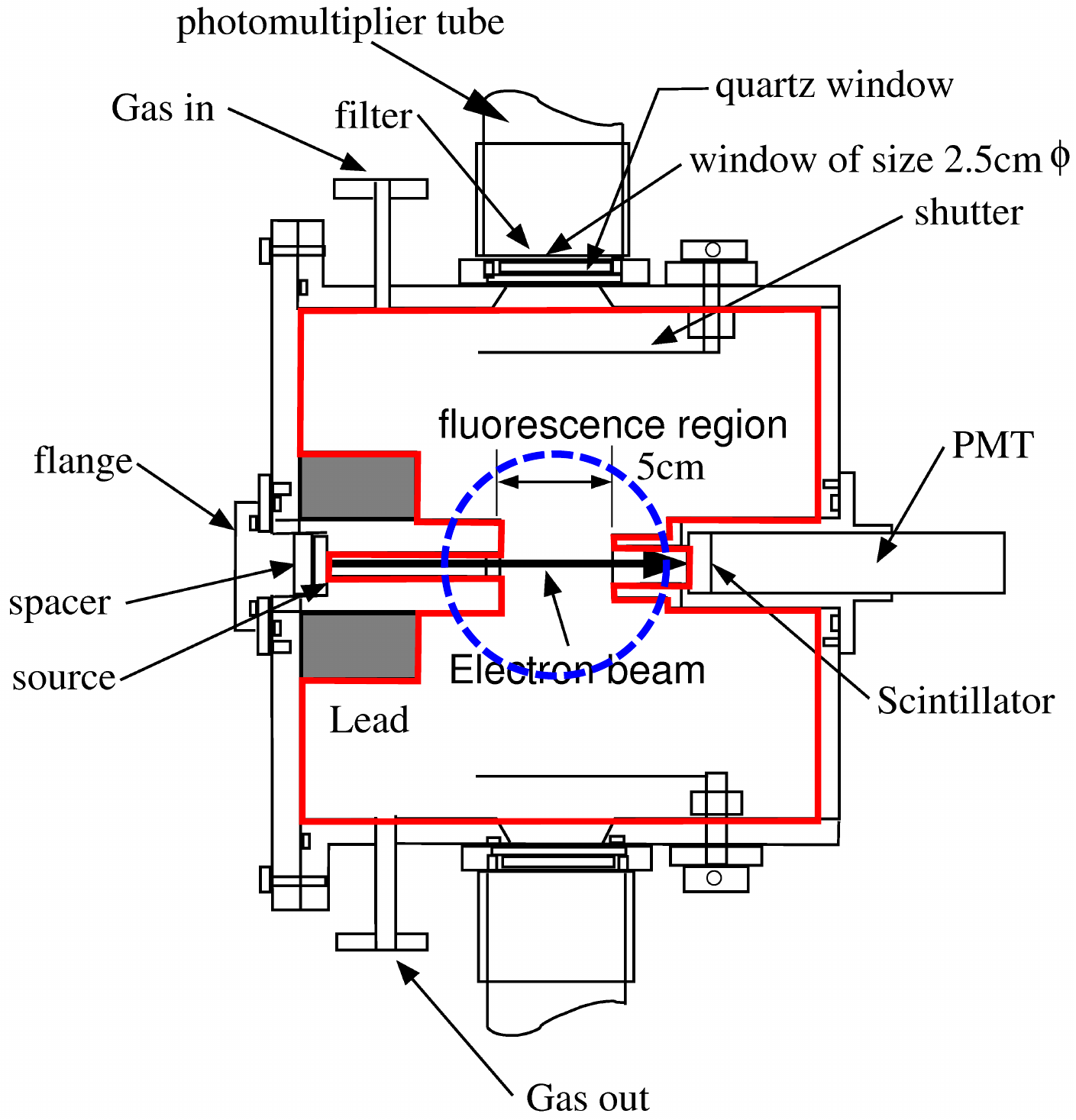}
\centering \caption{Schematic drawing of the chamber of Nagano's experiment (taken from~\cite{Nagano_03}). Electrons from a
$^{90}$Sr source are beamed and detected by a scintillator counter. Between the collimator and the scintillator enclosure there
is a 5~cm gap in the field of view of the photon detectors. The thick contour represents the geometry implemented in our
simulation. The energy deposition obtained from our simulation including these geometrical details is fully compatible with the
prediction for a sphere of 5~cm radius (circle).}\label{fig:Nagano_chamber}
\end{figure}

For Nagano's setup~\cite{Nagano_03,Nagano_04}, we made a detailed simulation in order to determine the effect of several
approximations applied in this experiment and then evaluate a correction to their results. The collisional chamber of this
experiment is shown in figure~\ref{fig:Nagano_chamber}, where the geometry implemented in this work is emphasized. They used the
electron-photon coincidence technique, so this feature has been also implemented in our simulation. Electrons fire a
scintillator at a distance of about 16~cm from the $^{90}$Sr source. However, only the 5~cm gap between the collimator and the
scintillator enclosure is in the field of view of the photon detectors. In this setup the observation volume corresponds to the
chamber volume excluding the collimation channel and the scintillator enclosure.

\begin{figure}
\includegraphics[width=\linewidth]{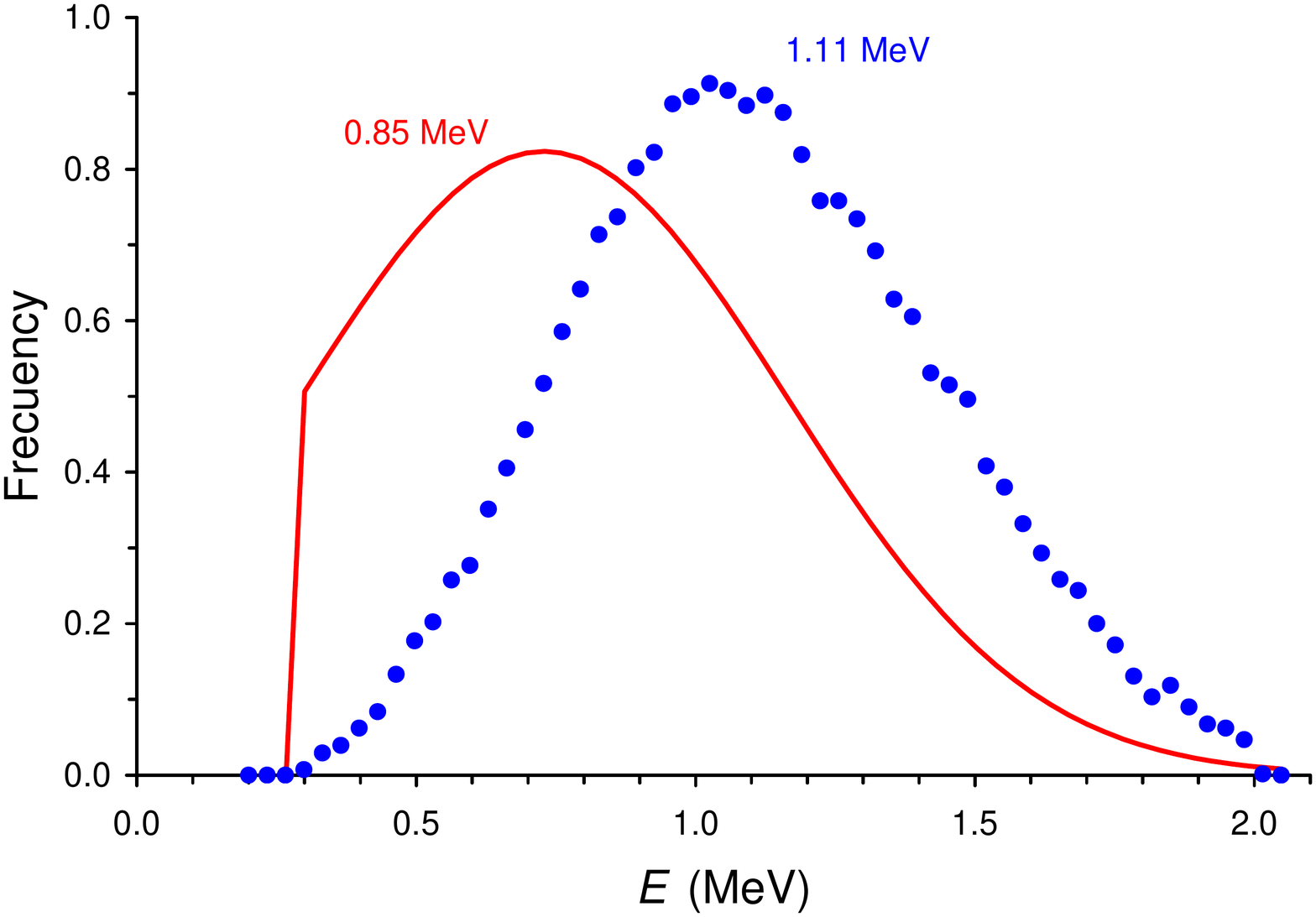}
\centering \caption{Energy spectrum used as input to the simulation of Nagano's experiment (solid line), with a mean energy of
0.85~MeV. According to our simulation, the energy distribution of electrons reaching the scintillator (symbols) is biased to
larger energies, resulting in a mean value of about 1.11~MeV at atmospheric pressure. Both spectra are normalized to the same
area.} \label{fig:Nagano_spectrum}
\end{figure}

As input to the simulation, we have used a truncated gaussian fit (solid line in figure~\ref{fig:Nagano_spectrum}) to the
measured $^{90}$Sr spectrum of~\cite{Nagano_03}, with mean energy of 0.85~MeV, end point at about 2~MeV and a threshold of
0.3~MeV due to the material covering the radioactive source. According to our simulation, most beam electrons are scattered away
at atmospheric pressure and so are not detected. This occurs preferably for low-energy primaries, and therefore the spectrum of
contributing electrons is shifted to higher energies, with a mean value of 1.11~MeV at atmospheric pressure (see
figure~\ref{fig:Nagano_spectrum}).

The approximations made by Nagano~\etal are the following. Firstly, they assumed that all the fluorescence light is emitted from
the beam itself. However, as explained above, a non-negligible fraction of fluorescence is produced by secondary electrons at
some distance, and thus, the average geometrical acceptance $\Omega$ may be different from that which would be obtained if all
the light were generated in the beam $\Omega_{\rm beam}$. Secondly, they calculated the $\varepsilon$ value assuming the photons
are emitted in a length equal to the gap distance $\Delta x_{\rm gap}=5$~cm. However, due to the fluctuating trajectories of
beam electrons, the mean path length $\Delta x$ should be somewhat larger. Finally, for the calculation of the fluorescence
yield from~(\ref{eq:Y}), they assumed that the energy deposited in the observation volume equals the collisional energy loss
$({\rm d}E/{\rm d}x)_{\rm loss}$ at 0.85~MeV as given by the Bethe-Bloch formula. As discussed above, the deposited energy is
expected to be smaller than the total energy loss.

Therefore, the fluorescence yield value $Y_{\rm Nag}$ reported by Nagano~\etal, should be corrected by three factors accounting
for the above described approximations

\begin{equation}\label{eq:Nagano}
Y=Y_{\rm Nag}\frac{\Omega_{\rm beam}}{\Omega}\frac{\Delta x_{\rm gap}}{\Delta x}\frac{\left({\rm d}E/{\rm d}x\right)_{\rm
loss}}{\langle{\rm d}E/{\rm d}x\rangle_{\rm dep}}\,.
\end{equation}

From our detailed simulation, we have obtained the average geometrical acceptance relative to that assumed
in~\cite{Nagano_03,Nagano_04}, that is, the $\Omega/\Omega_{\rm beam}$ ratio in the above expression. For this calculation, we
have taken into account that some regions in the observation volume are hidden to the optical system. We have found that this
correction to the efficiency is very small, as expected, resulting $\Omega$ to be about 1\% lower than $\Omega_{\rm beam}$ at
atmospheric pressure. The correction is even smaller at lower pressure. On the other hand, according to our simulation, the mean
path length of contributing electrons at atmospheric pressure is about 1\% longer than the gap length. Therefore, these small
corrections happen to compensate each other in equation~(\ref{eq:Nagano}) for this experiment. In fact, the Nagano's results on
$\varepsilon$ (photons/meter) turn out to be unaffected by these approximations.

The main correction to the fluorescence yield in equation~(\ref{eq:Nagano}) arises from the approximation in the energy
deposition. From our simulation, we have obtained the average energy deposited per electron and unit path length $\langle{\rm
d}E/{\rm d}x\rangle_{\rm dep}$ as the ratio between the integrated energy deposition and the total path length of beam electrons
within the observation volume. Simulation data were accumulated up to reach a statistical uncertainty of less than 1\% in the
$\langle{\rm d}E/{\rm d}x\rangle_{\rm dep}$ value. As a result of this correction, the fluorescence yield at atmospheric
pressure from~\cite{Nagano_03,Nagano_04} should be increased by 7\%.

The result of the deposited energy from our simulation of the Nagano's experiment is fully compatible with the predictions from
the generic simulation for a simple geometry, assuming a sphere of radius $R=5$~cm (see figure~\ref{fig:Nagano_chamber}) and an
electron energy of around 1~MeV. As already mentioned, deposited energy is very weakly dependent on~$R$. In addition, at these
energies around the minimum of the Bethe-Bloch formula, the result is nearly independent of~$E$. Therefore, for experiments
similar to that of Nagano, the generic simulation may provide reliable results on the energy deposited in the observation
volume.

\subsection{The AirLight experiment}
\label{sec:AirLight}

The experimental technique used in AirLight~\cite{AirLight} is similar to the one of Nagano, but with a collisional chamber
larger by about a factor of two. Electrons from a $^{90}$Sr source are collimated to form a narrow beam of 5~mm diameter.
Between the collimator and the electron detector (a scintillator of 4~cm diameter) there is a 10~cm gap within the field of view
of seven photomultipliers. In~\cite{AirLight} a detailed simulation of the experiment is carried out using GEANT4, including
electron scattering by elements of the chamber (e.g., collimator walls, scintillator). Both beam electrons and secondary ones
are thoroughly tracked allowing the authors the determination of the geometrical acceptance as well as the energy deposited in
the observation volume.

\begin{figure}
\includegraphics[width=\linewidth]{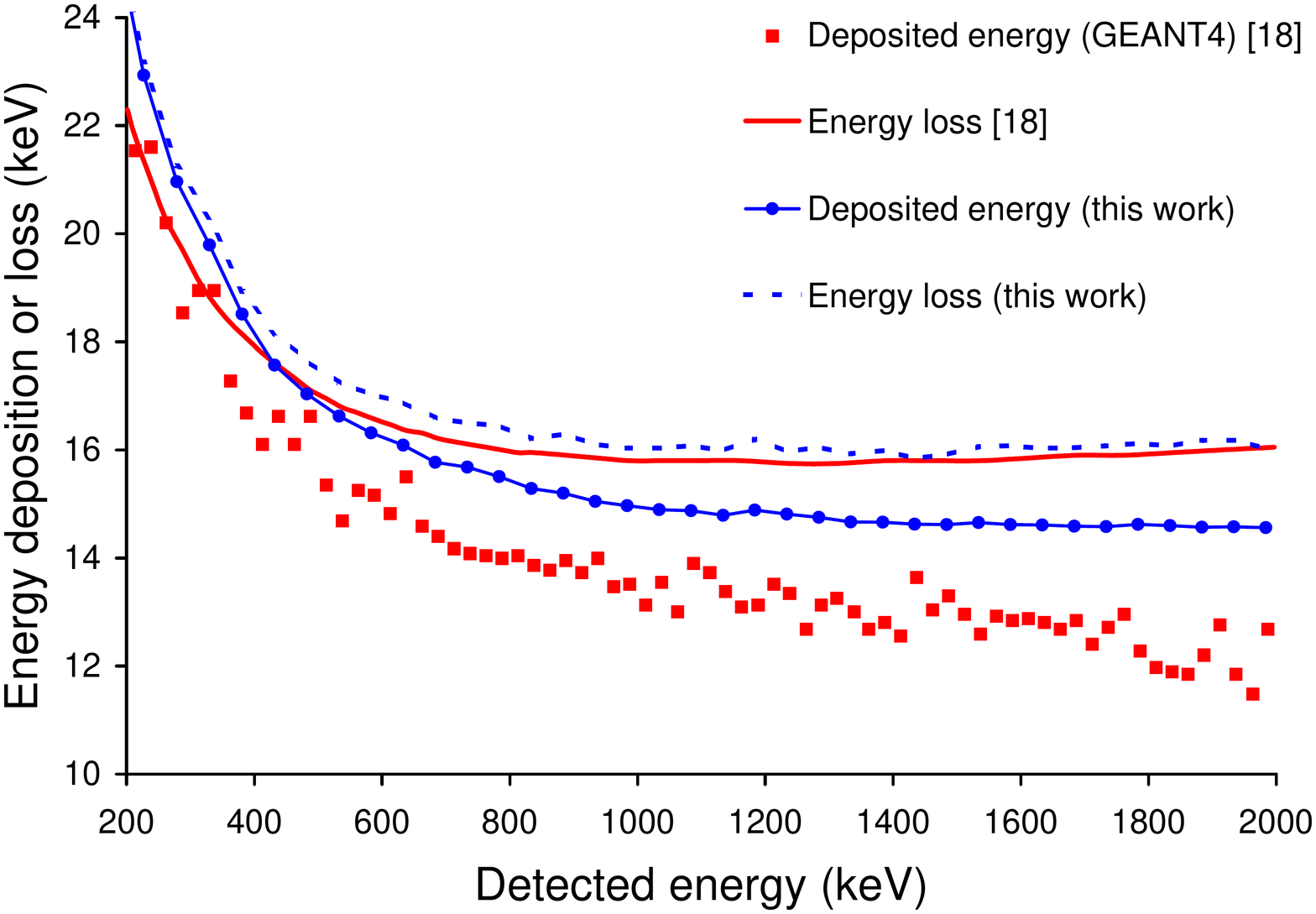}
\centering \caption{Integrated energy deposition in the chamber of the AirLight experiment as a function of the electron energy
detected in the scintillator in the 200 - 2000~keV range~\cite{AirLight}. Results at 800~hPa obtained by~AirLight using GEANT4
(squares) are compared with ours (circles connected by lines). Integrated energy loss from our simulation (dotted line) is
compared with that calculated by AirLight (solid line). See text for more details.}\label{fig:AirLight}
\end{figure}

For this experiment we have carried out a simulation including primary propagation and the geometrical details of the setup.
Electron scattering by other elements of the chamber apart from air is not included. The aim of our simulation has been the
calculation of the integrated energy deposition as a function of the electron energy to be compared with the results found by
AirLight. While at low pressure we have obtained results in fair agreement with those reported in~\cite{AirLight} (within 5\% at
50~hPa), at atmospheric pressure our simulation predicts a larger energy deposition. For instance at 800~hPa, deviations range
from about 5\% at low electron energy up to 20\% at 2000~keV (see figure~\ref{fig:AirLight}). The origin of this discrepancy is
still unclear. It should be indicated that electron scattering by components of the collision chamber (included in their
simulation, but not in ours) would result indeed in an increase of the energy deposition, as pointed out in~\cite{AirLight}.

From our simulation we have also calculated the integrated value of the collisional energy loss. At atmospheric pressure we
obtain results somewhat larger than those reported by~\cite{AirLight}, in particular at low electron energy (see
figure~\ref{fig:AirLight}). The energy loss computed by AirLight seems to have been calculated as the product of the Bethe-Bloch
stopping power $({\rm d}E/{\rm d}x)_{\rm loss}$~\cite{Seltzer} times the 10~cm gap length, and thus, assuming that the stopping
power is constant along the electron path. However, as already mentioned for Nagano's experiment, due to the fluctuating
trajectories of primaries, their mean path length is expected to be somewhat larger than the gap length. Also, at low electron
energy, the average stopping power should be larger than the initial value, because primaries lose a significant fraction of
their energy in their path.

Obviously, the energy deposition must be smaller than the energy loss, and their difference is expected to diminish at low $E$
values. This behavior is found in our simulation as well as in that of~\cite{AirLight}. However, at 800~hPa, the energy loss
from~\cite{AirLight} is smaller than our prediction of the energy deposition for electron energies below 400~keV
(figure~\ref{fig:AirLight}) showing an evident inconsistency between both calculations.

Assuming a mean electron energy of 1~MeV, our detailed simulation predicts an average deposited energy around 10\% higher than
that of~\cite{AirLight} at 800~hPa, and thus, if our results were correct, their fluorescence yield value at this pressure
should be reduced in principle by the same factor. However, AirLight reports $Y^0_{vv'}$ results (together with a set of
quenching parameters) by averaging measurements carried out at several pressures\footnote{AirLight found a slight, although
unexpected, pressure dependence of $Y^0_{vv'}$.}. In fact the average $Y^0_{vv'}$ values are nearly equal to those obtained at
400~hPa. At this pressure and an electron energy of 1~MeV, the difference between our energy deposition and that of AirLight is
about 7\%. Therefore, according to our simulation, results on fluorescence yield from~\cite{AirLight} should be reduced by an
overall correction factor of 7\% irrespective of pressure.

\subsection{The FLASH experiment}
\label{sec:FLASH}

In this setup~\cite{FLASH_08}, an intense beam of 28.5~GeV electrons from a linear accelerator crosses a cylindrical collision
chamber, 25~cm in length and 15~cm in diameter. The beam is coaxial to a pair of 1.6~cm diameter cylinders with a 1.67~cm gap
between them, which defines the beam path length as seen by two PMTs. Each detector is mounted at the end of a long channel
perpendicular to the beam axis and provided with baffles to suppress unwanted scattered light. This channel terminates, at 45~cm
from the beam, in a 1.2~cm diameter window that forms the entrance pupil of the optical system. In the opposite side, there is a
shorter channel holding a LED used to monitor the PMT gain stability, being the extreme of this channel at about 27~cm from the
beam (72~cm from the window of the PMT). The absolute calibration of the optical system was carried out by comparison with the
measurement of Rayleigh-scattered light from a narrow beam of a nitrogen laser. As pointed out above, the FLASH Collaboration
used an EGS4 simulation to obtain the mean energy deposited per electron in the observation volume of their
experiment~\cite{FLASH_08}. Their simulation also provided the average geometrical acceptance for fluorescence detection
relative to that for the calibration laser beam.

We have carried out a detailed simulation of this experiment in order to evaluate the energy deposited in the chamber as well as
the average geometrical acceptance for comparison with their results. In our algorithm, the observation volume has been
approximated by a rectangular prism of dimensions 1.60~cm$\times 1.67$~cm$\times 72$~cm.

\begin{figure}
\includegraphics[width=\linewidth]{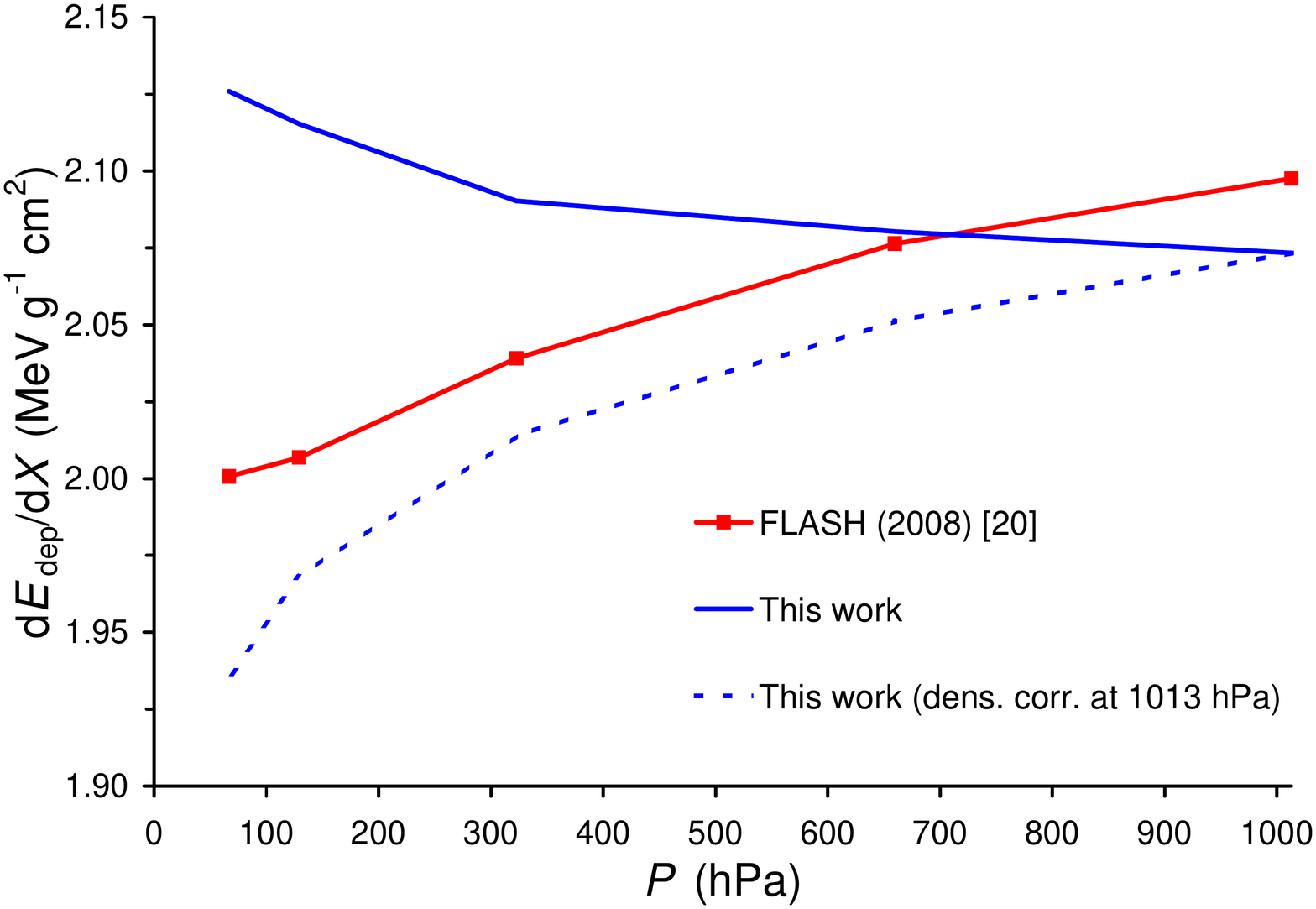}
\centering \caption{Energy deposited per electron and unit mass thickness as a function of pressure for the FLASH
experiment~\cite{FLASH_08}. Results from~\cite{FLASH_08} (squares connected by lines) and from our detailed simulation (solid
line) are in good agreement at atmospheric pressure, although the pressure dependency is different. For comparison, results from
our simulation applying the density correction at 1013~hPa in the whole pressure range are also shown (broken
line).}\label{fig:FLASH_08}
\end{figure}

The deposited energy is not explicitly given in~\cite{FLASH_08}, but it can be inferred from the ratio between the reported
results on $\varepsilon$ and $Y$ at pressures ranging from 67 to 1013~hPa. In figure~\ref{fig:FLASH_08}, the resulting values
are shown as a function of pressure together with those from our detailed simulation\footnote{The results of our detailed
simulation are in excellent agreement (within 1\%) with that obtained from the generic simulation for a sphere of 1.67~cm
radius.}, where statistical uncertainties were reduced to less than 0.3\%. At high pressures ($\gtrsim 500$~hPa), both results
are in good agreement (within 2\%). However, the pressure dependency is different, resulting in a deviation of 6\% at 67~hPa.
This discrepancy could be due to a different treatment of the density correction, which at such a high electron energy is very
important, lowering the inelastic cross sections by about 40\% at atmospheric pressure. As a simple test, we have recalculated
the deposited energy using in our simulation the density correction for 1013~hPa in the whole pressure range. The resulting
energy deposition, also presented in figure~\ref{fig:FLASH_08}, shows a behavior similar to that of~\cite{FLASH_08}.

In regard with the geometrical acceptance, FLASH~\cite{FLASH_08} found that the efficiency of their optical system for
fluorescence detection is $(3.2\pm 0.25)$\% lower than that for the calibration laser beam, due to the wide-spread energy
deposition from the electron beam. However, our simulation predicts that this effect is negligible in the whole pressure range
for this setup.

Notice that according to our simulation both corrections, i.e., deposited energy and optical efficiency, nearly cancel and
therefore the fluorescence yield at 1013~hPa from~\cite{FLASH_08} would be almost unaffected (see table~\ref{tab:comparison} in
section~\ref{sec:results}).

\subsection{The MACFLY experiment}
\label{sec:MACFLY}

In this experiment~\cite{MACFLY}, the collision chamber is a cylinder 28.8~cm long and 15~cm in diameter with a mirrored inner
surface in order to increase the geometrical acceptance. They used the coincidence technique, with individual electrons either
from a linear accelerator (20 and 50~GeV) or from a $^{90}$Sr source of mean energy 1.5~MeV. As mentioned above, MACFLY
calculated the energy deposited per electron and per unit mass thickness as a function of $E$ using GEANT4 (blue dotted line in
figure~\ref{fig:Edep_comparison}).

We have carried out a detailed simulation to obtain the energy deposition. Primaries colliding the walls of the chamber or being
stopped inside, and so not reaching the trigger system, are rejected in the simulation. The average energy deposited per
electron and unit mass thickness has been obtained from the ratio between the integrated energy deposition and the total path
length of triggering primaries. Statistical uncertainties are reduced below 0.5\%. The deposited energy obtained from our
simulation is slightly lower (2\%) than the value reported by MACFLY at 1.5~MeV and larger by about 6\% at 20 and 50~GeV, and
therefore, according to our calculations the fluorescence yield of MACFLY should be increased by 2\% at 1.5~MeV and decreased by
6\% at 20 and 50~GeV.

The above described detailed simulation is in good agreement (better than 2\%) with the generic simulation for $R=10$~cm at
energies above 0.5~MeV (see figure~\ref{fig:Edep_comparison}). Below 0.5~MeV, primary electrons lose a significant fraction of
their energy in the chamber (e.g., electrons with energy lower than 0.2~MeV are fully stopped), and consequently the
$\langle{\rm d}E/{\rm d}x\rangle_{\rm dep}$ values resulting from our detailed simulation are somewhat larger ($\sim 5\%$) than
those obtained from the generic one. However this effect cannot justify the very large discrepancies with the MACFLY results at
low electron energy shown in figure~\ref{fig:Edep_comparison}.

\subsection{Other experiments}
\label{sec:others}

Apart from those discussed in the above sections other well-know experiments have been carried out, in particular those of
Kakimoto~\etal~\cite{Kakimoto} and Lefeuvre~\etal~\cite{Lefeuvre}. For these we have not performed a detailed simulation but we
will use the result of our generic simulation.

In the Kakimoto's experiment electrons with a mean energy of 1.4~MeV from a $^{90}$Sr source as well as those from an electron
synchrotron with energies of 300, 650 and 1000~MeV were used. Fluorescence light was produced and observed in a 10~cm gap. For
the determination of the fluorescence yield, they assumed full energy deposition in the beam axis. According to our generic
simulation, the deposited energy inside a sphere of 10~cm radius is about 6\%, 25\%, 28\% and 29\% lower than the energy loss
for 1.4, 300, 650 and 1000~MeV electrons, respectively. Therefore, the fluorescence yield values at the above energies should be
increased correspondingly by these factors. Results obtained from applying these corrections to measurements of Kakimoto~\etal
are consistent with those obtained in~\cite{Arqueros_Moriond} using an alternative procedure.

In the experiment of Lefeuvre~\etal~\cite{Lefeuvre}, electrons with mean energies of 1.1~MeV (or 1.5~MeV when applied a higher
energy threshold to the electron detector) from a very intense $^{90}$Sr source (370~MBq) were used. In order to shield the
fluorescence detector from the large amount of X-radiation produced at such a high electron rate, both the radioactive source
and the scintillator counter were placed inside a solid lead cylinder, 10~cm in diameter. Electrons trajectories were confined
within a cone dug in the lead and coaxial to the cylinder, being the $^{90}$Sr source at the apex and the scintillator on the
base. A 4~cm diameter hole was bored through the lead cylinder, allowing fluorescence to be measured from both sides. In order
to obtain the average geometrical acceptance and the mean path length of electrons, they performed a GEANT Monte Carlo
simulation including electron scattering by the lead walls. They found that this feature has an important role in this
experiment, since half of primaries reach the scintillator after hitting the cone walls. From their simulation, they also found
that a very small fraction of secondary electrons have enough energy to escape the observation region at atmospheric pressure.
In particular, they obtained that only 1\% of secondaries have energies above 5~keV, and 33\% of them are lost in the lead
before having produced any fluorescence. From these simulation results, they inferred that the deposited energy nearly equals
the total energy loss and a minor correction of 0.4\% was applied. However, as discussed above, a significant fraction of energy
is deposited by these few energetic secondaries, and thus, the correction is expected to be larger.

From our generic simulation assuming a sphere of 4~cm radius for Lefeuvre's experiment, we have obtained that the deposited
energy is 9\% and 10\% lower than the total energy loss at mean electron energies of 1.1 and 1.5~MeV, respectively. However,
electron scattering by the lead shield of~\cite{Lefeuvre} is not included in our simulation, which assumes that secondaries
reaching the walls are absorbed, and thus, it may underestimate the deposited energy in this case. We have evaluated this effect
in order to estimate a correction to our simulation results for this experiment. Using CASINO2.42~\cite{CASINO}, we have found
that, for a $60^{\circ}$ incident angle and energies in the 1~keV~-~1~MeV range, about 60\% of electrons are backscattered by
lead, having lost almost half of their energy on the average. Assuming that backscattered electrons deposit all their remainder
energy in the air volume, about 30\% of the energy of electrons reaching the lead walls is recovered, and thus, should be added
to the total energy deposition. As a consequence, the fluorescence yield values at 1.1 and 1.5~MeV from~\cite{Lefeuvre} should
be increased by about 7\% according to our calculations.

\section{Wavelength interval} \label{sec:wavelength_interval}

The wavelength normalization can be carried out if the relative intensities of the molecular bands within the experimental
interval are known. As is well known, relative intensities of bands sharing the same upper level are given by the relative
Einstein coefficients $B_{vv'}=A_{vv'}/\sum_{v'}{A_{vv'}}$. As described below fluorescence intensities of bands with different
upper levels can be also calculated from molecular parameters.

In the absence of other population mechanisms\footnote{As indicated in section~\ref{sec:units}, the effect of vibrational
relaxation of the 2P upper levels can be considered as included in the $P'_{v}$ quenching parameters.} apart from direct
excitation of the vibrational level $v$ from the ground state~X, the band intensities of a given system are proportional to the
excitation cross sections of the corresponding upper levels. At high electron energy (i.e., in the Born-Bethe high-energy
limit), these cross sections are known to be approximately proportional to the Franck-Condon factors $q_{{\rm X}\rightarrow v}$,
which correspond to the probability ratios for optical transitions. It has been found~\cite{Arqueros_NJP} that this
approximation can be also applied in our case, in spite of the fact that most fluorescence light is generated by low-energy
secondaries. Under this approximation, for a given band system, the relative intensities $I_{vv'}$ with respect to a reference
one $I_{00}$ (i.e., the most intense band of the system, located at 337~nm for the 2P system and at 391~nm for the 1N system)
are given by

\begin{equation}\label{eq:relative_intensities}
\frac{I_{vv'}}{I_{00}}=\frac{q_{X\rightarrow v}B_{vv'}}{q_{X\rightarrow 0}B_{00}}\,\frac{1+P/P'_0}{1+P/P'_v}\,.
\end{equation}

At atmospheric conditions, where $P\gg P'_v$ can be assumed, the above expression becomes

\begin{equation}\label{eq:relative_intensities_limit}
\frac{I_{vv'}}{I_{00}}\approx\frac{q_{X\rightarrow v}B_{vv'}P'_v}{q_{X\rightarrow 0}B_{00}P'_0}\,,
\end{equation}
where any pressure or temperature dependence is cancelled, since the same approximate $\sqrt{T}$ growing behavior can be assumed
for all $P'_v$ parameters.

Relationship~(\ref{eq:relative_intensities_limit}), and hence the applicability of the Franck-Condon factors, has been verified
using experimental data on relative intensities. In figure~2 of ref.~\cite{Arqueros_NJP} we plotted the relative intensities
measured by AIRFLY~\cite{AIRFLY_P} against $q_{X\rightarrow v}B_{vv'}P'_v$ for the 2P system, using the tabulated values of
$q_{X\rightarrow v}$ and $B_{vv'}$ from~\cite{Gilmore,Laux} and the experimental characteristic pressures for dry air reported
in~\cite{AIRFLY_P}. A remarkable linearity was found, in particular for the bands with upper vibrational level $v=0,\,1$,
contributing to almost 90\% of the total 2P fluorescence. A similar analysis was previously performed in~\cite{Blanco} using the
relative intensities and $P'_v$ values measured by Nagano~\etal~\cite{Nagano_04}. Although linearity was also observed,
significant fluctuations were found very likely due to much larger experimental uncertainties than those of~\cite{AIRFLY_P}.

Expression~(\ref{eq:relative_intensities_limit}) can be extended for molecular bands belonging to different systems including
the ratio between the respective effective cross sections $\sigma_{\rm syst}^{\rm eff}$, defined as $\sigma_{vv'}^{\rm
eff}=\sigma_{\rm syst}^{\rm eff}\, q_{X\rightarrow v}B_{vv'}$~\cite{Arqueros_NJP}. Unfortunately, whereas the Franck-Condon
factors and the Einstein coefficients are known very precisely, the available data on $\sigma_{\rm syst}^{\rm eff}$ and $P'_v$
are less reliable. Effective cross sections depend on molecular parameters with significant uncertainties and are geometry
dependent, and the available experimental data on characteristic pressures show large discrepancies. However, as already pointed
out in~\cite{Arqueros_NJP}, the relative intensities measured by different authors, which are proportional to $\sigma_{\rm
syst}^{\rm eff}P'_v$, are in good agreement. For instance, for the $I_{337}/I_{391}$ ratio at atmospheric pressure, values of
$3.0\pm 0.7$, $3.7\pm 0.4$, $3.4\pm 0.3$ and $3.57\pm 0.13$ result from data from~\cite{Davidson}, \cite{Nagano_03},
\cite{Nagano_04} and~\cite{AIRFLY_P}, respectively. Even a very consistent 3.5 value was measured by Bunner~\cite{Bunner_PhD}
for 4~MeV $\alpha$-particles as well. Therefore, the value of the $\sigma_{\rm syst}^{\rm eff}P'_v$ products  from experimental
relative intensities is very reliable.

In order to compare absolute results on fluorescence yield from different authors, we will reduce values measured in a given
wavelength interval $\Delta\lambda$ to the yield at the 337~nm band. Experimental results are then normalized multiplying by the
ratio between the relative intensity of the 337~nm band and the sum of all the bands within $\Delta\lambda$. This
$I_{337}/I_{\Delta\lambda}$ ratio can be thus written in terms of the empirically reliable $I_{337}/I_{391}$ ratio as

\begin{equation}\label{eq:ratio_337_interval}
\frac{I_{337}}{I_{\Delta\lambda}}=\left[\sum_{\Delta\lambda}\frac{I_{vv'}^{\rm 2P}}{I_{337}}+
\frac{I_{391}}{I_{337}}\sum_{\Delta\lambda}\frac{I_{vv'}^{\rm 1N}}{I_{391}}\right]^{-1}\,,
\end{equation}
where relative intensities at atmospheric conditions follow~(\ref{eq:relative_intensities_limit}) separately for each band
system. Note that, due to the 2P dominance, even an error of a factor 2 in the $I_{337}/I_{391}$ ratio results in an uncertainty
below 10\% in $I_{337}/I_{\Delta\lambda}$ for the spectral ranges of the experiments considered here.

For this wavelength reduction, we use accurate experimental relative intensities from~\cite{AIRFLY_P} in the 284 - 429~nm
spectral range, including some bands of the Gaydon-Herman (GH) system~\cite{AIRFLY_P} (1.6\% total contribution). For instance,
for the whole wavelength interval of~\cite{AIRFLY_P} a ratio $I_{337}/I_{[284-429]}=0.2569\pm 0.0022$ is obtained, where the
uncertainties on the relative intensities are treated as uncorrelated as the authors suggest. Note that this value is slightly
lower than those listed in table~1 of ref.~\cite{Arqueros_NJP}, where intensities of bands of the GH system were neglected. In
addition, equations~(\ref{eq:relative_intensities_limit}) and~(\ref{eq:ratio_337_interval}) allow the determination of
$I_{337}/I_{\Delta\lambda}$ for a wavelength interval including some of the 2P and 1N weak bands outside the spectral range
of~\cite{AIRFLY_P}.

\section[Results]{Results of the fluorescence yield \\comparison}
\label{sec:results}

In table~\ref{tab:comparison} absolute fluorescence yield values at atmospheric pressure reported by several authors (first
column) are put in comparison. The wavelength interval, pressure, temperature and electron energy at which absolute values are
reported are shown in columns~2~-~5. The experimental results, in units of either photons/MeV or photons/m, and the
corresponding uncertainties provided by the authors are quoted in columns~6 and~7. In order to get values normalized to that of
the 337~nm band we have applied the procedure shown in section~\ref{sec:wavelength_interval}. The corresponding normalization
factors are displayed in column~8. The last column shows the fluorescence yield for the 337~nm band in units of photons/MeV. For
an easy comparison, all results are given for 800~hPa and 293~K (dry air). For this normalization we have used
expression~(\ref{eq:Y0}) assuming $P\gg P'_{337}$ and $P'_{337}\propto\sqrt{T}$. The numbers in the left side of this column
correspond to the $Y_{337}$ value obtained using the energy deposition and geometrical factors assumed (or calculated) by the
authors. Finally, in the right side (in bold) we show the results we have obtained applying the various corrections discussed in
section~\ref{sec:detailed_simulation}.

\begin{sidewaystable}
%\begin{table*}[t!]
\caption{Comparison of absolute values of fluorescence yield from several experiments. Experimental results as given by the
authors are quoted in column~6. The last column shows the fluorescence yield in units of photons/MeV resulting from the
normalization to 337~nm, 800~hPa and 293~K (dry air) using the assumptions/calculations of the authors (left side) or the
results from our simulations (right side). See text for details.}
\centering\small%
\begin{tabular}{*{9}{c}}
\addlinespace\toprule
Experiment                       &    $\Delta\lambda$ (nm)    &       $P$ (hPa)       &        $T$ (K)       &     $E$ (MeV)    & Experimental result &        Error           & $I_{337}/I_{\Delta\lambda}$ & $Y_{337}$ (ph/MeV) \\
\midrule\midrule
\multirow{5}{*}{Kakimoto~$[14]$} &             337            &           800         &          288         &        1.4       &      5.7 ph/MeV     &         10\%           &           1                 & 5.8 / \textbf{6.1} \\
\cmidrule{2-9}
                                 & \multirow{4}{*}{300 - 400} & \multirow{4}{*}{1013} & \multirow{4}{*}{288} &        1.4       &      3.3 ph/m       & \multirow{4}{*}{10\%}  &    \multirow{4}{*}{0.279}   & 5.7 / \textbf{6.0} \\
                                 &                            &                       &                      &      300         &      4.9 ph/m       &                        &                             & 5.6 / \textbf{7.0} \\
                                 &                            &                       &                      &      650         &      4.4 ph/m       &                        &                             & 4.8 / \textbf{6.1} \\
                                 &                            &                       &                      &     1000         &      5.0 ph/m       &                        &                             & 5.4 / \textbf{6.9} \\
\midrule
Nagano~$[16]$                    &             337            &          1013         &          293         &        0.85      &      1.021 ph/m     &         13\%           &           1                 & 6.4 / \textbf{6.8} \\
\midrule
\multirow{2}{*}{Lefeuvre~$[17]$} & \multirow{2}{*}{300 - 430} & \multirow{2}{*}{1005} & \multirow{2}{*}{296} &        1.1       &      3.95 ph/m      &  \multirow{2}{*}{5\%}  &    \multirow{2}{*}{0.262}   & 6.5 / \textbf{7.0} \\
                                 &                            &                       &                      &        1.5       &      4.34 ph/m      &                        &                             & 7.1 / \textbf{7.7} \\
\midrule
\multirow{3}{*}{MACFLY~$[13]$}   & \multirow{3}{*}{290 - 440} & \multirow{3}{*}{1013} & \multirow{3}{*}{296} &        1.5       &     17.0 ph/MeV     & \multirow{3}{*}{13\%}  &    \multirow{3}{*}{0.255}   & 5.5 / \textbf{5.6} \\
                                 &                            &                       &                      &  $20\cdot 10^3$  &     17.4 ph/MeV     &                        &                             & 5.6 / \textbf{5.3} \\
                                 &                            &                       &                      &  $50\cdot 10^3$  &     18.2 ph/MeV     &                        &                             & 5.9 / \textbf{5.5} \\
\midrule
FLASH~$[20]$                     &          300 - 420         &          1013         &          304         & $28.5\cdot 10^3$ &     20.8 ph/MeV     &          7.5\%         &         0.272               & 7.0 / \textbf{6.9} \\
\midrule
AirLight$^a~[18]$                &             337            &           -           &           -          &      0.2 - 2     & $Y^0=384$ ph/MeV    &           16\%         &           1                 & 7.4 / \textbf{6.9} \\
\midrule
AIRFLY$^b~[21]$                  &             337            &           993         &          291         &       350        &      4.12 ph/MeV    &          -             &           1                 & 5.1 / \textbf{ - } \\
\bottomrule\addlinespace%
\end{tabular}

\raggedright\footnotesize%
$^a$The listed experimental result is the fluorescence yield at null pressure reported by the authors which, together with their
measured quenching parameters, allows the evaluation of the fluorescence yield at any atmospheric condition.

$^b$Preliminary result. \label{tab:comparison}
\end{sidewaystable}
%\end{table*}

As explained in section~\ref{sec:simple_geometry}, our simulation gives us both deposited energy and fluorescence emission and
thus we can provide a theoretical result of the fluorescence yield~\cite{5th_FW_Arqueros}. The updated calculations of this work
lead us to a $Y^0_{337}$ value of 390~ph/MeV, nearly independent of the electron energy above 0.1~MeV. This theoretical result
in combination with the characteristic pressure for the 337~nm band in dry air~\cite{AIRFLY_P} gives rise to a $Y_{337}$ value
of 7.5~ph/MeV at 800~hPa and 293~K, which is close to the experimental ones shown in the table. However, as pointed out above,
uncertainties in the simulated fluorescence emission are large, and thus, this theoretical result cannot be considered as a
reference one.

\section{Conclusions}
\label{sec:conclusions}

A direct comparison of the absolute measurements of the air fluorescence yield available in the literature is not possible since
very often the reported data correspond to different wavelength intervals and the magnitude is expressed in different units. The
conversion factors for such comparison depend on the particular features of the experimental setup. In this work we have
described a procedure for the normalization of the various experimental results to a common unit and wavelength interval.

In the first place, fluorescence yields have been reduced to that corresponding to the most intense band at 337~nm using
accurate experimental data on intensity ratios in full agreement with theoretical relationships. Absolute values can be
expressed in photons/MeV units as far as the energy deposited in the field of view of the setup is known. In this work we have
used an upgraded simulation algorithm to calculate the energy deposition of electrons in a wide energy range. A generic
algorithm has provided results as a function of the pressure and the overall size of the collision chamber. In addition, for
several experiments, we have performed a detailed simulation (in full agreement with the predictions of the generic one)
including in some cases the calculation of geometrical factors. The results of these simulations have been compared with those
reported by the authors of the corresponding experiments and eventual disagreements have been discussed.

As a result of this work a table of the absolute fluorescence yields normalized to 800~hPa and 293~K have been presented.
According to our simulation, a non-negligible correction has to be applied to those results obtained under the assumption that
the energy lost by the primary electrons is fully deposited in the field of view of the optical system (i.e.,
Kakimoto~\etal~\cite{Kakimoto}, Nagano~\etal~\cite{Nagano_03,Nagano_04}) and Lefeuvre~\etal~\cite{Lefeuvre}). In particular the
Nagano's fluorescence yield should be increased by about 7\%. Possible implications of these corrections to the $Y$ values in
the determination of the end of the energy spectrum of cosmic rays have been discussed recently in~\cite{Nagano_NJP}.

On the other hand, our calculations of energy deposition are in general in reasonable agreement with those from experiments that
carried out a detailed simulation of their setup. Even though, in some cases we have found non-negligible discrepancies and
correction factors have been proposed. Experiments claiming high accuracy should be very careful in the evaluation of the error
contribution from the deposited energy.

From the above table, and assuming the corrections proposed in this work, most fluorescence yield measurements at those
reference conditions are contained within the 6~-~7~ph/MeV interval. Notice, however that the fluorescence yield reported by
MACFLY, as well the preliminary result of AIRFLY, are smaller by more than 10\%.

\section*{Acknowledgements}

This work has been supported by the Spanish Ministerio de Ciencia e Innovacion (FPA2006-12184-C02-01, FPA2009-07772 and
CONSOLIDER CPAN CSD2007-42) and ``Comunidad de Madrid" (Ref.: 910600). J.~Rosado acknowledges a PhD grant from ``Universidad
Complutense de Madrid".

\end{document}